\documentclass[reprint,aps,pra,superscriptaddress,amsmath,amssymb
]{revtex4-2}
\usepackage[utf8]{inputenc}
\usepackage{graphicx}
\usepackage{dcolumn}
\usepackage{bm}
\usepackage{mathrsfs}
\usepackage{amsfonts}
\usepackage{color}
\usepackage{xcolor}
\usepackage{braket}
\usepackage[english]{babel}
\usepackage{url}
\definecolor{darkblue}{rgb}{0,0,0.5}
\usepackage{hyperref}
\hypersetup{
colorlinks=true,
linkcolor=black,
filecolor=blue,
citecolor=darkblue,  
urlcolor=black,
}

\usepackage[absolute]{textpos}

\def\cI{\mathcal{I}}

\begin{document}

\title{Non-Gaussian photonic state engineering with the quantum frequency processor}

\author{Andrew J. Pizzimenti}
\email{ajpizzimenti@email.arizona.edu}
\affiliation{Quantum Information Science Section, Oak Ridge National Laboratory, Oak Ridge, Tennessee 37831, USA}
\affiliation{James C. Wyant College of Optical Sciences, University of Arizona, Tucson, Arizona 85721, USA}
\author{Joseph M. Lukens}
\affiliation{Quantum Information Science Section, Oak Ridge National Laboratory, Oak Ridge, Tennessee 37831, USA}
\author{Hsuan-Hao Lu}
\affiliation{Elmore Family School of Electrical and Computer Engineering and Purdue Quantum Science and Engineering Institute, Purdue University, West Lafayette, Indiana 47907, USA}
\author{Nicholas A. Peters}
\affiliation{Quantum Information Science Section, Oak Ridge National Laboratory, Oak Ridge, Tennessee 37831, USA}
\author{Saikat Guha}
\author{Christos N. Gagatsos}
\affiliation{James C. Wyant College of Optical Sciences, University of Arizona, Tucson, Arizona 85721, USA}

\begin{abstract}
Non-Gaussian quantum states of light are critical resources for optical quantum information processing, but methods to generate them efficiently remain challenging to implement. Here we introduce a generic approach for non-Gaussian state production from input states populating discrete frequency bins. Based on controllable unitary operations with a quantum frequency processor, followed by photon-number-resolved detection of ancilla modes, our method combines recent developments in both frequency-based quantum information and non-Gaussian state preparation. Leveraging and refining the $K$-function representation of quantum states in the coherent basis, we develop a theoretical model amenable to numerical optimization and, as specific examples, design quantum frequency processor circuits for the production of Schr\"{o}dinger cat states, exploring the performance tradeoffs for several combinations of ancilla modes and circuit depth. Our scheme provides a valuable framework for producing complex quantum states in frequency bins, paving the way for single-spatial-mode, fiber-optic-compatible non-Gaussian resources.
\end{abstract}

\maketitle

\section{introduction}\label{sec:introduction}
The distinction between discrete-variable (DV) and continuous-variable (CV) encodings offers a valuable lens through which to classify and understand photonic quantum information processing systems. Based on true (or approximate) finite-dimensional Hilbert spaces, DV optical designs are typically associated with qubits encoded in photons that are manipulated and subsequently measured with single-photon detectors~\cite{Knill2001,Kok2007}. On the other hand, the infinite-dimensional Hilbert spaces of CV quantum information exploit collective photonic excitations (such as coherent or squeezed states) and homodyne/heterodyne detection with local oscillators as fundamental resources~\cite{Lloyd1999, Bartlett2002, Braunstein2005, Weedbrook2012}. From a technical side, the DV/CV divide can prove quite stark, and significant differences appear theoretically as well: for example, security proofs for CV quantum key distribution have generally proven much more challenging to establish due to the infinite dimensionality involved~\cite{Diamanti2015, Ghorai2019}.

Yet this dichotomy is far from absolute, with features of particular quantum information processing approaches blurring the CV/DV distinction entirely. At the implementation level, many DV photonic systems utilize subspaces taken from a larger, intrinsically continuous Hilbert space---time~\cite{Brendel1999, Marcikic2002, Humphreys2013, Islam2017} and frequency bins~\cite{Lukens2017, Kues2019, Lu2019c} forming representative examples of relevance to the present work. 
In an even more direct fashion, in encodings such as the Gottesman--Kitaev--Preskill (GKP) qubit~\cite{Gottesman2001, Fukui2018, Tzitrin2020}, the \emph{logical} quantum information is discrete, but the \emph{encoding} occupies the full continuous Hilbert space. Here the CV aspects are not incidental features of the chosen Hilbert space; rather, they prove critical to the paradigm itself, providing the foundation for measuring and correcting continuous errors on the logical qubit state.

The potential of error-corrected photonic quantum information processing with GKP qubits makes them an appealing direction for research. But producing such states---and non-Gaussian CV states more generally---is an extremely challenging endeavor, with proof-of-principle GKP realizations so far limited to non-photonic platforms~\cite{Fluhmann2019, Campagne2020}. The discovery and analysis of Gaussian boson sampling (GBS)~\cite{Hamilton2017, Kruse2019}, however, has provided a valuable framework for preparing non-Gaussian optical states \cite{mattia2021,mattia2020,Su2019,Gagatsos2019}, based on earlier important works on the universality of Gaussian states and partial post-selection \cite{Cerf2005,Janszky2000,Boas2001}. Also straddling the interface between CV and DV---in that it leverages both CV fields and single-photon detection---GBS circuits can in principle produce arbitrary non-Gaussian states through ancilla modes and postselection on particular detection patterns, analogous to the probabilistic gates of linear-optical quantum computation (LOQC) in the DV paradigm~\cite{Knill2001, Kok2007}. The design~\cite{Gagatsos2019, mattia2020, Su2019, Sabapathy2019, Quesada2019, Tzitrin2020} and implementation~\cite{Paesani2019, Zhong2019, Arrazola2021}  of GBS-type circuits for non-Gaussian state preparation have so far focused on the path degree of freedom (DoF), a natural choice given its long history in optics and well-known unitary decomposition procedure~\cite{Reck1994, Clements2016}. But other DoFs offer promise as well. As the focus of the present work, the frequency-bin DoF enjoys several attractive features for scalable photonic quantum information processing, including wavelength parallelizability, compatibility with single-mode optical fiber, and CV state production with resonant parametric oscillators, both free-space~\cite{Pysher2011, Chen2014} and integrated~\cite{Dutt2015, Vaidya2020}.

A major challenge of non-Gaussian state production with frequency-bin encoding, however, is the realization of arbitrary unitary operations. Recent work on the quantum frequency processor (QFP)~\cite{Lu2019c} in the LOQC mold has made significant strides to this end; based on alternating application of electro-optic phase modulators (EOMs) and pulse shapers, the QFP can in principle synthesize any unitary frequency-bin operation in a scalable fashion. Following the original proposal~\cite{Lukens2017}, the QFP has been demonstrated experimentally on both single-~\cite{Lu2018a, Lu2020b} and two-photon~\cite{Lu2018b, Lu2019a} states. Yet apart from a classical communications example using quadrature-encoded data~\cite{Lu2020a}, the research focus has been entirely within the DV paradigm, so that the opportunities and limitations of applying the QFP to CV---and hybrid DV/CV---systems remain uncharted.

In this work, we develop and simulate a complete model for non-Gaussian frequency-bin state engineering on the QFP. Leveraging and expanding on the $K$-function formalism of Ref.~\cite{Gagatsos2019}, we describe a resource-efficient method for computing the output of a QFP excited by Gaussian inputs and measured with photon-number-resolving (PNR) detectors applied to a subset of frequency modes. As examples of this general approach, we design QFP circuits intended to produce Schr\"{o}dinger cat states in one undetected bin and explore the impact of the number of components and ancilla modes on circuit performance, according to a cost function which balances both state fidelity and success probability. Our approach furnishes a general framework for non-Gaussian state production in frequency-bin quantum systems, offering a springboard for the design of practical experimental systems.

\section{Mathematical Background of our Approach}\label{sec:Mathematical Background of our Approach}
For modeling our proposed system, we use a representation of Gaussian states in the coherent basis according to the $K$-function formalism introduced in Ref.~\cite{Gagatsos2019}. Here we briefly review the results in Ref.~\cite{Gagatsos2019} and then evolve those to further worked-out formulas. Among other things, in Ref.~\cite{Gagatsos2019} it was proven that any $N$-mode pure Gaussian state $|\Psi\rangle$ with covariance matrix (CM) $V$ and displacement vector $\vec{x}_\beta$ can be written in the coherent basis $|\vec{\alpha}\rangle$ as
\begin{eqnarray}
\label{eq:PsiWithDisplacements}|\Psi\rangle = \int d^{2N} \vec{x}_\alpha\ K(\vec{x}_\alpha) |\vec{\alpha} \rangle,
\end{eqnarray}
where
\begin{eqnarray}
\label{eq:Kfunction} K(\vec{x}_{\alpha})&=&\frac{e^{-\frac{1}{2} (\vec{x}_{\alpha}-\vec{x}_{\beta})^T \mathcal{B} (\vec{x}_{\alpha}-\vec{x}_{\beta})+\frac{1}{2} \vec{x}_\alpha^T \mathcal{Y} \vec{x}_\beta }}{(2\pi)^N(\det \Gamma)^{1/4}},
\end{eqnarray}
with $\Gamma=V+I/2$,
\begin{eqnarray}
\label{eq:Bmatrix}	\mathcal{B}&=&\frac{1}{2}
	\begin{pmatrix}
	A + \frac{i}{2}\left(C+C^T\right) & C - \frac{i}{2}\left(A-B\right) \\
	C^T - \frac{i}{2}\left(A-B\right) & B - \frac{i}{2}\left(C+C^T\right)
	\end{pmatrix},\\
	\mathcal{Y}&=&
	\begin{pmatrix}
	0 & iI \\
	-iI& 0
	\end{pmatrix},
\end{eqnarray}
where $A=A^T$, $B=B^T$, and $C$ are defined as the blocks of $\Gamma^{-1}$ as follows:
\begin{eqnarray}
\label{eq:GammaInv}	\Gamma^{-1}=
	\begin{pmatrix}
	A & C\\
	C^T & B
	\end{pmatrix}.
\end{eqnarray}
Note that we have simplified the expressions compared to Ref.~\cite{Gagatsos2019}.
We note that since the CM $V$ is symmetric, $\Gamma$ and $\Gamma^{-1}$ are also symmetric.
We work with the convention $\hbar=1$ (therefore the CM of vacuum is $I/2$) and consider the $qqpp$ representation where vectors are defined as  ${\vec{x}_{\alpha}^T=(\vec{q}_{\alpha}^T,\vec{p}_{\alpha}^T)}$ 
with $\vec{q}^T_{\alpha}=(q_{\alpha_1},\ldots,q_{\alpha_N})$ and $\vec{p}_{\alpha}^T=(p_{\alpha_1},\ldots,p_{\alpha_N})$ the canonical position and momentum vectors.
The volume element for integration is then defined as $d^{2N} \vec{x}_{\alpha}=dq_{\alpha_1}\ldots dq_{\alpha_N} dp_{\alpha_1}\ldots dp_{\alpha_N}$, and $\alpha_i=(q_{\alpha_i}+i p_{\alpha_i})/\sqrt{2}$.

The coherent basis representation is a valuable tool for working on photon-subtraction-based or, more generally, partial PNR detection schemes aimed at engineering Gaussian states into desired non-Gaussian states. Photon subtraction can be modelled either (i) as a beamsplitter whose two input ports are fed with the $i$th mode of $|\Psi\rangle$ and vacuum $|0\rangle$, respectively, followed by PNR detection on the lower output port; or (ii) simply by acting the annihilation operator $\hat{a}_i$, where the index $i$ refers to the mode, on $|\Psi\rangle$. Therefore, the photon subtraction operator will act only on the basis vectors of the state, i.e., coherent states in this instance. The action of beamsplitters or annihilation operators on coherent states is straightforward, making this basis particularly efficient for analytical or numerical evaluation. 
The situation is similar for partial PNR detection on a Gaussian state written as a coherent state expansion; 
the projection of a coherent state on a Fock state is the well known expression $\langle n|\alpha \rangle = \exp(-|\alpha|^2/2)\alpha^n/\sqrt{n!}$.

In Ref.~\cite{Gagatsos2019} it was shown that the probability of a length-$N$ PNR pattern 
for an $N$-mode Gaussian state $|\Psi\rangle$ with zero displacements, i.e., $\vec{x}_\beta=0$ in Eq.~\eqref{eq:PsiWithDisplacements}, is given by
\begin{eqnarray}
 \nonumber   P_{n_1 \ldots n_N}&=&|\langle n_1 \ldots n_N| \Psi\rangle|^2 \\
\label{eq:GBSprob} &=&  \frac{1}{\det \mathcal{H} \sqrt{\det\Gamma} \prod\limits_{i=1}^{N} n_i!2^{n_i}}\big|\mathcal{I}_{n_1 \ldots n_N}\big|^2,
\end{eqnarray}
where
\begin{eqnarray}
\label{eq:IN}	\mathcal{I}_{n_1 \ldots n_N} &=&\int d^{2N}\vec{x}_\alpha R(\vec{x}_{\alpha})  \prod\limits_{i=1}^{N}(q_{\alpha_i}+i p_{\alpha_i})^{n_i},\\
\label{eq:Rdistr}	R(\vec{x}_{\alpha}) &=&  \frac{\sqrt{\det \mathcal{H}}}{(2\pi)^{N}}  e^{-\frac{1}{2}\vec{x}_\alpha^T \mathcal{H} \vec{x}_\alpha},
\end{eqnarray}
and $\mathcal{H}=\mathcal{B}+I/2$. Equation \eqref{eq:IN} can be rewritten as
\begin{eqnarray}
\label{eq:Hafnian}	\mathcal{I}_{n_1 \ldots n_N} = \left\{
\begin{array}{ll}
0&\Sigma=\textrm{odd},\\
\textrm{Hf}\left(\sigma\right)&\Sigma= \textrm{even},
\end{array}
\right.
\end{eqnarray}
where $\Sigma=\sum_{i=1}^{N}n_i$, $\textrm{Hf}\left(\sigma\right)$ is the hafnian (often specifically called the ``loop hafnian'' in the literature~\cite{Su2019}) of the matrix $\sigma$ with elements $\sigma_{ij}=\langle s_i s_j \rangle$, where $1\leq i,j\leq \Sigma$ and  $s_i=q_{\alpha_i}+ip_{\alpha_i}$. The hafnian in Eq. \eqref{eq:Hafnian} represents the mean value $\langle s_1^{n_1}\ldots s_N^{n_N} \rangle$ under the Gaussian distribution of Eq.~\eqref{eq:Rdistr}.

In this work, we will derive the explicit relation of the matrix $\sigma$ to the matrix $\mathcal{H}^{-1}$ and consequently to the matrix $\Gamma^{-1}$ which describes the Gaussian state just before partial PNR detection.  This enables more efficient computation of the output detection probabilities and the Fock coefficients of any produced non-Gaussian state, for a given input Gaussian state. We also simplify further the expressions. The following subsections summarize new simplifications, observations, and new results which improve on Eqs. (\ref{eq:GammaInv}--\ref{eq:Rdistr}).

\subsection{The determinant and inverse of $\Gamma$}\label{sec:Gamma}
The matrix $\Gamma$ is defined as $\Gamma=V+I/2$, where $V$ is the CM and $I$ the identity matrix. Since $V$ corresponds to a pure Gaussian state, it can be written as $V=S_p V_0 S_p^T$, where $S_p$ is an orthogonal symplectic matrix for a general passive transformation (beamsplitters and phase rotations, but not squeezers) and $V_0$ is the CM for a product of $N$ single mode squeezed vacuum states, i.e., the diagonal matrix
\begin{eqnarray}
V_0=\frac{1}{2}\textrm{diag} \left( e^{2 r_1},\ldots,e^{2 r_N}, e^{-2 r_1},\ldots, e^{-2 r_N}\right),
\end{eqnarray}
where $r_1,\ldots,r_N$ are the real and positive squeezing parameters for each of the $N$ single-mode squeezed vacuum states (note that the phase of the squeezing has been absorbed into the orthogonal symplectic transformation $S_p$).

We have the following relation,
\begin{eqnarray}
\det\Gamma&=&\det\left[S_p\left(V_0+\frac{I}{2}\right)S_p^T\right]\\
&=&\det S_p \det\left(V_0+\frac{I}{2}\right) \det S_p^T,
\end{eqnarray}
from which we write
\begin{eqnarray}
\label{eq:detGamma1}\det\Gamma= \det\left(V_0+\frac{I}{2}\right)
\end{eqnarray}
since $\det S_p=\det S_p^T=1$ as both $S_p$ and $S_p^T$ are symplectic matrices.
The right hand side of Eq. \eqref{eq:detGamma1} is the determinant of a diagonal matrix from which we find
\begin{eqnarray}
\label{eq:detGamma2}\det\Gamma = \prod_{i=1}^N \cosh^2 r_i.
\end{eqnarray}
Therefore, Eq. \eqref{eq:GBSprob} is rewritten as
\begin{eqnarray}
\label{eq:GBSprob2} P_{n_1 \ldots n_N}=\frac{\big|\mathcal{I}_{n_1\ldots n_N}\big|^2}{\det \mathcal{H} \prod\limits_{i=1}^{N} n_i!2^{n_i} \cosh r_i}.
\end{eqnarray}
In the case where the input squeezing is the same among all single mode squeezed vacuum states, i.e. $r_1=\ldots=r_N=r$, Eq. \eqref{eq:detGamma2} reduces to $\det\Gamma=\cosh^{2N}r$.

Now let us simplify Eq. \eqref{eq:GammaInv}. We can write $\Gamma=S_p (V_0+I/2)S_p^T$, and since $S_p^{T^{-1}}=S_p$ is a symplectic orthogonal matrix we have
\begin{eqnarray}
\label{eq:GammaInv2} \Gamma^{-1} = S_p \left(V_0+\frac{1}{2}\right)^{-1} S_p^T.
\end{eqnarray}
The symplectic orthogonal matrix $S_p$ has the following block matrix structure and properties:
\begin{eqnarray}
\label{eq:Mp}S_p &=& \begin{pmatrix}
S_A & S_B\\
-S_B & S_A
\end{pmatrix}\\
\label{eq:MpConstr1}&& S_A^T S_B = S_B^T S_A,\\
\label{eq:MpConstr2}&& S_A S_B^T = S_B S_A^T,\\
\label{eq:MpConstr3}&& S_A^T S_A+S_B^T S_B = I,\\
\label{eq:MpConstr4}&& S_A S_A^T+S_B S_B^T = I.
\end{eqnarray}
Moreover, since $V_0$ is diagonal we can write
\begin{eqnarray}
\label{eq:invGamma_0}
\left(V_0+\frac{1}{2}\right)^{-1} = I+\begin{pmatrix}
-T & 0\\
0 & T
\end{pmatrix},
\end{eqnarray}
where $T=\textrm{diag}\left(\tanh r_1,\ldots,\tanh r_N\right)$. In virtue of Eqs. \eqref{eq:GammaInv2}, \eqref{eq:Mp}, and \eqref{eq:MpConstr2}, we find that in Eq. \eqref{eq:GammaInv} 
\begin{eqnarray}
\label{eq:blockA} A &=&-S_A T S_A^T + S_B T S_B^T,\\
\label{eq:blockC} C&=&C^T= S_A T S_B^T+S_B T S_A^T,\\
&& A+B= 2I.
\end{eqnarray}
Therefore, in the most general case possible, Eq. \eqref{eq:GammaInv} is simplified to
\begin{eqnarray}
\label{eq:GammaInv3} \Gamma^{-1} = \begin{pmatrix}
A & C \\
C & 2 I -A,
\end{pmatrix}
\end{eqnarray}
where $A$ and $C$ are given in Eqs. \eqref{eq:blockA} and \eqref{eq:blockC}, respectively, as functions of the passive symplectic transformation $S_p$ and the input squeezing parameters.

Consequently, matrix $\mathcal{B}$ of Eq. \eqref{eq:Bmatrix} simplifies to
\begin{eqnarray}
\label{eq:Bmatrix2}\mathcal{B}=\frac{1}{2}\begin{pmatrix}
A+ i C & C-i(A-I) \\
C-i(A-I) & 2I-A-iC
\end{pmatrix}.
\end{eqnarray}
\subsection{The determinant and inverse of $\mathcal{H}$}\label{sec:H}
The matrix $\mathcal{H}$ appearing in Eq.~\eqref{eq:Rdistr} is defined as 
\begin{eqnarray}
\label{eq:Hmatrix}\mathcal{H}=\mathcal{B}+I/2. 
\end{eqnarray}
We find it easier if we transform as $\tilde{\mathcal{H}}= W^\dagger\mathcal{H}W$ using the unitary matrix $W$ 
defined as
\begin{eqnarray}
\label{eq:Wmatrix}W=\frac{1}{\sqrt{2}}\begin{pmatrix}
I & I \\
-i I & i I
\end{pmatrix}.
\end{eqnarray}
Utilizing Eqs. \eqref{eq:Bmatrix2}, \eqref{eq:Hmatrix}, and \eqref{eq:Wmatrix} we find
\begin{eqnarray}
\label{eq:Hmatrix2}\tilde{\mathcal{H}}= \begin{pmatrix}
I & A-I +i C\\
0 & I
\end{pmatrix},
\end{eqnarray}
from which we see that $\det \tilde{\mathcal{H}}=\det I=1$. Since $|\det W|^2=1$, we have $\det \tilde{\mathcal{H}}=\det\mathcal{H}$ and conclude that
\begin{eqnarray}
\label{eq:detH}\det\mathcal{H}=1.
\end{eqnarray}
Therefore, Eqs. \eqref{eq:Rdistr} and  \eqref{eq:GBSprob2} are further simplified to
\begin{eqnarray}
\label{eq:GBSprob3} P_{n_1 \ldots n_N}&=&\frac{\big|\mathcal{I}_{n_1\ldots n_N}\big|^2}{ \prod\limits_{i=1}^{N} n_i!2^{n_i} \cosh r_i},\\
\label{eq:Rdistr2}	R(\vec{x}_{\alpha}) &=&  \frac{1}{(2\pi)^{N}}  e^{-\frac{1}{2}\vec{x}_\alpha^T \mathcal{H} \vec{x}_\alpha}.
\end{eqnarray}
Let us derive a convenient expression for $\mathcal{H}^{-1}$. Again, we work with $\tilde{\mathcal{H}}$ and observe that
\begin{eqnarray}
\label{eq:HmatrixTildeInv}\tilde{\mathcal{H}}^{-1}= \begin{pmatrix}
I & -(A-I +i C)\\
0 & I
\end{pmatrix}
\end{eqnarray}
is indeed the inverse of $\tilde{\mathcal{H}}$, i.e., it satisfies $\tilde{\mathcal{H}} \tilde{\mathcal{H}}^{-1}=I$. Since $\tilde{\mathcal{H}}= W^\dagger\mathcal{H}W$ we find that $\mathcal{H}^{-1} = W \tilde{\mathcal{H}}^{-1} W^\dagger$ and finally
\begin{eqnarray}
\label{eq:Hinv}\mathcal{H}^{-1}=\frac{1}{2}
\begin{pmatrix}
3 I -A-i C & i (A-I+i C)\\
i (A-I+i C) & I+A+ i C
\end{pmatrix}.
\end{eqnarray}
Therefore, using Eqs. \eqref{eq:blockA}, \eqref{eq:blockC}, and \eqref{eq:Hinv}, any given passive symplectic transformation $S_p$, and input squeezing parameters, one can readily write $\mathcal{H}^{-1}$---the importance of which will become apparent in the next subsections.
\subsection{The relation of matrix $\sigma$ to matrix $\mathcal{H}^{-1}$}\label{sec:FandHinv}

Making use of Eq.~\eqref{eq:Rdistr2}, we can express the matrix elements of $\sigma$ as
\begin{eqnarray}
\nonumber \sigma_{ij}&=&\langle \left(q_{\alpha_i}+i p_{\alpha_i}\right)\left(q_{\alpha_j}+i p_{\alpha_j}\right)\rangle=\\
\nonumber && \frac{1}{(2\pi)^N}\int d^{2 N} \vec{x}_{\alpha} \exp \left(-\frac{1}{2} \vec{x}_{\alpha}^{T} \mathcal{H} \vec{x}_{\alpha}\right)\\
\nonumber &&\times\left(q_{\alpha_{i}}+i p_{\alpha_{i}}\right)\left(q_{\alpha_{j}}+i p_{\alpha_{j}}\right)=\\
\label{eq:FandHinv}&& \left.\frac{d}{d \lambda_{i}} \frac{d}{d \lambda_{j}} \exp \left(\frac{1}{2} \vec{\Lambda}^{T} \mathcal{H}^{-1} \vec{\Lambda}\right)\right|_{\vec{\Lambda}=\overrightarrow{0}},
\end{eqnarray}
where $\vec{\Lambda}^T=(\vec{\lambda}^T, i \vec{\lambda}^T)$ is a $2N$-dimensional vector with $\vec{\lambda}^T=\left(\lambda_1,\ldots,\lambda_N\right)$ a real $N$-dimensional vector. Viewing $\frac{1}{2} \vec{\Lambda}^{T} \mathcal{H}^{-1} \vec{\Lambda}$ in the exponential of the right hand side of Eq. \eqref{eq:FandHinv} as a polynomial in $\lambda_i$,  Eq. \eqref{eq:FandHinv} is equal to the coefficient of $\lambda_i\lambda_j$. This way, we can write
\begin{eqnarray}
\label{eq:FandHinv2}\sigma_{ij} = 2 (\mathcal{H}^{-1}_{ij}-\mathcal{H}^{-1}_{i+N\ j+N}).
\end{eqnarray}
From the covariance matrix $V$, one can find matrix $\Gamma^{-1}$ and therefore matrix $\sigma$ using Eqs. \eqref{eq:Hinv} and \eqref{eq:FandHinv2}, which is required in the calculation in Eq. \eqref{eq:Hafnian}.

The Gaussian moment problem of Eq. \eqref{eq:IN} represents a hafnian calculation and is related to the Gaussian boson sampling paradigm \cite{Hamilton2017}. When the indices $i,j$ are equal this corresponds to a loop, i.e., matching an object with itself. Therefore, it is typically referred to as a loop hafnian. 
\subsection{Occurrence probability of any produced state}\label{sec:Probability}
\begin{figure}
		\includegraphics[width=\columnwidth]{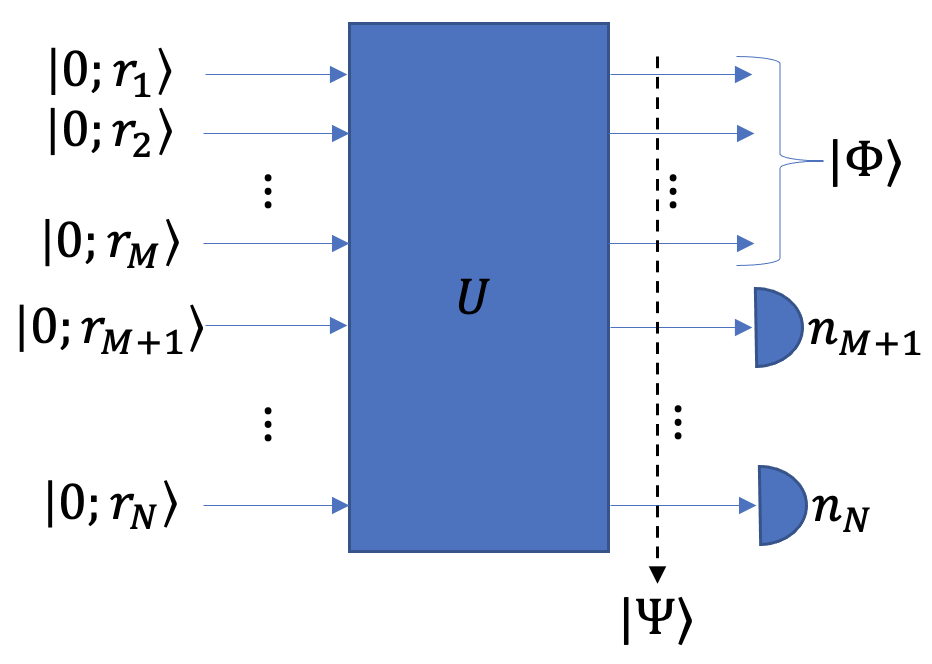}
		\caption{Concept of heralding an $M$-mode state $\ket{\Phi}$ from $N$ single-mode, zero-displacement squeezed resource states and $N\times N$ unitary operation $U$. 
		Partial PNR detection on the $N-M$ lower modes produces a non-Gaussian state on the undetected $M$ modes. 
		}
		\label{fig:GenericScheme}
\end{figure}
Equation \eqref{eq:GBSprob3} is the probability of finding $n_i$ photons in each one of the $i=1,\ldots,N$ modes. If we wish to engineer the $N$-mode Gaussian state into an $M$-mode ($M<N$) non-Gaussian one as in Fig. \ref{fig:GenericScheme}, we leave $M$ modes undetected; without loss of generality we assume the undetected modes are the $M$ upper modes. The probability of the PNR pattern $(n_{M+1}, \ldots , n_N)$ on the lower detected modes is precisely the probability $P_{n_{M+1}, \ldots ,n_N}$ of producing the corresponding non-Gaussian state. This probability is
\begin{eqnarray}
\label{eq:P}P\equiv P_{n_{M+1},\ldots,n_N} = \sum_{n_1,\ldots,n_M=0}^{\infty} P_{n_1,\ldots,n_N}.
\end{eqnarray}
For numerical simulations, the above sum must be truncated to a finite upper limit, which should be chosen with care to ensure that it encompasses all Fock coefficients of nonnegligible probability. This condition can be verified in practice by successively increasing the limits and observing no change to $P$.
\subsection{Fock expansion coefficients of the produced state}
The non-Gaussian state $|\Phi\rangle$ on the $M$ undetected modes (see Fig.~\ref{fig:GenericScheme}), can be written as a partial projection on Fock states of the detected modes:

\begin{eqnarray}
\nonumber|\Phi\rangle &=&\frac{1}{\sqrt{P}} \langle n_{M+1}\ldots n_N | \Psi \rangle\\
\nonumber &=& \frac{1}{\sqrt{P}} \sum_{n_1,\ldots, n_M=0}^\infty \langle n_1\ldots n_M n_{M+1}\ldots n_N | \Psi \rangle |n_1\ldots n_M\rangle\\
\label{eq:phi}&=& \sum_{n_1,\ldots, n_M=0}^\infty c_{n_1\ldots n_M} |n_1\ldots n_M\rangle ,
\end{eqnarray}
where $P$ is given in Eq. \eqref{eq:P}, $|\Psi\rangle$ is the $N$-mode Gaussian state just before partial PNR detection (i.e., the output Gaussian state), and  $c_{n_1 \ldots n_M}=\langle n_1\ldots n_M|\Phi\rangle$ are the Fock expansion coefficients of the heralded state $|\Phi\rangle$.

 Using Eqs. \eqref{eq:IN} and \eqref{eq:phi} we find
\begin{eqnarray}
\label{eq:FockCoef}c_{n_1 \ldots n_M} = \frac{\cI_{n_1 \ldots n_M n_{M+1}\ldots n_N}}{\sqrt{P} \prod\limits_{i=1}^{N} \sqrt{n_i!2^{n_i} \cosh r_i}},
\end{eqnarray}
where the numerator is given by Eq. \eqref{eq:IN}. Therefore, for any given partial PNR pattern $(n_{M+1},\ldots,n_N)$ one can compute the Fock expansion coefficients of the produced state $|\Phi\rangle$, which can be benchmarked against a target non-Gaussian state $\ket{\Phi_t}$ through direct comparison of Fock coefficients or collectively through fidelity $\mathcal{F}=|\braket{\Phi_t|\Phi}|^2$.

\subsection{Summarizing Comments}
Let us close Sec.~\ref{sec:Mathematical Background of our Approach} with three remarks.
First, we note that our formalism provides an approach to computing Gaussian states in the Fock basis complementary to that of Refs.~\cite{Su2019, Quesada2019}.
By incorporating the reduced dimensionality of a pure state directly, our approach requires calculation of fewer expansion coefficients to fully characterize the output, in the case of pure state evolution; for example, for a photon cutoff of $n_c$, a pure single-mode output state is described by $n_c+1$ complex coefficients $c_{n_K}$ [Eq.~\eqref{eq:Fockcoeff_N}], while a mixed single-mode state formulation would require $(n_c+1)(n_c+2)/2$, after accounting for hermiticity. 


Second, extra care is required when dealing with loop hafnians. Let us give an example. Say that one wants to calculate $\langle s_1^2 s_2 s_3 \rangle$.  
To apply Wick's formula one has to rewrite the mean value as containing four different objects, i.e., $\langle s_1^2 s_2 s_3 \rangle=\langle g_1 g_2 g_3 g_4 \rangle$. Wick's formula gives the perfect matchings as $\langle g_1 g_2 g_3 g_4 \rangle=\langle g_1 g_2\rangle \langle g_3 g_4\rangle+\langle g_1 g_3\rangle \langle g_2 g_4\rangle+\langle g_1 g_4\rangle \langle g_2 g_3\rangle$, and then we substitute back $g_1=g_2=s_1,\ g_3=s_2,\ \textrm{and}\ g_4=s_3$, which gives
$\langle s_1^2 s_2 s_3\rangle=\langle s_1^2\rangle \langle s_2 s_3\rangle+\langle s_1 s_2\rangle \langle s_1 s_3\rangle+\langle s_1 s_3\rangle \langle s_1 s_2\rangle$.

For the calculation of loop hafnians, we find it more efficient instead 
to work with the formula found in Ref.~\cite[Prop. 1, p. 547]{Kan2008} as it takes inherently into account the powers of $s_i$. For the convenience of the reader we give the formula (adjusted to our notation) which is the nonzero result of Eq. \eqref{eq:Hafnian}:
\begin{eqnarray}
\nonumber \textrm{Hf}(\sigma) &=& \frac{1}{\left(\frac{\Sigma}{2}\right)!} \sum_{\nu_1=0}^{n_1} \ldots \sum_{\nu_N=0}^{n_N} (-1)^{\nu_1+\ldots+\nu_N}\\
\label{eq:FasterWick} && \times \binom{n_1}{\nu_1}\ldots \binom{n_N}{\nu_N}\left(\frac{1}{2}\vec{h}^T \sigma \vec{h} \right)^{\frac{\Sigma}{2}}
\end{eqnarray}
where $\vec{h}^T=\left(n_1/2-\nu_1,\ldots,n_N/2-\nu_N\right)$, and $\nu_i,\ldots,\nu_N$ are silent indices, i.e., they are summed. 
Equations \eqref{eq:IN} and \eqref{eq:FasterWick} can be used directly in Eq. \eqref{eq:P} for the probability of finding any non-Gaussian state in the undetected modes and in Eq. \eqref{eq:FockCoef} for the Fock expansion coefficients of such a state.
This tailored expression for the loop hafnian was noted for its significant computational speed up in previous non-Gaussian state engineering work as well~\cite{Quesada2019}. Essentially, the improvement is obtained when the dominate bottleneck in Wick's formula stems from repeated factors (e.g., $s_1$ in the example above) rather than many non-repeated factors, (e.g., $s_2$ and $s_3$ in the example above). This is certainly the case in our work, where we consider many photons in the single undetected mode to fully characterize the post-selected state (up to $n_\mathrm{max}\sim 40$), with only a few PNR detectors (2 or 4).

Third and finally, the formulas above enable calculation of the coefficients $\braket{n_1 ... n_N|\Psi}$ for any diagonal input covariance matrix $V_0$ and passive symplectic mode transformation $S_p$---i.e., any covariance matrix for a pure Gaussian state---without numerical evaluation of a single matrix inverse  or determinant: these expressions have all been reduced to straightforward matrix or scalar operations in the above. This simplification has a profound impact on the efficiency of the numerical procedure in Sec.~\ref{sec:numOpt}, eliminating  time-consuming inverse calculations from the optimization loop. 


\section{Quantum frequency processor}\label{sec:QFP}
\begin{figure}[tb!]
    \centering
    \includegraphics[width=\columnwidth]{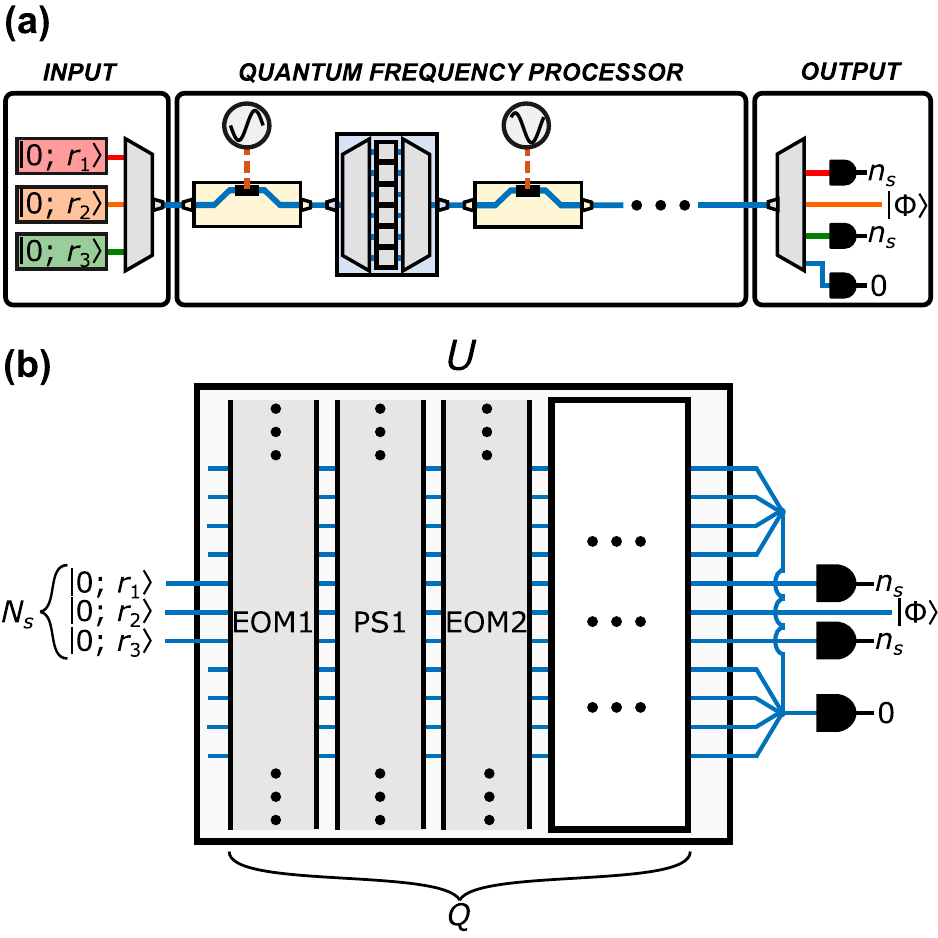}
    \caption{Setup explored for non-Gaussian state preparation with the QFP. The case of $N_s=3$ input squeezed modes is shown for concreteness.  (a) Hardware view. Squeezed states in distinct frequency modes traverse the sequence of EOMs and pulse shapers in the QFP. The condition for successful heralding is the detection of $n_s$ photons each in all but one of the central $N_s$ bins and zero photons in all adjacent bins.  The undetected mode is left in state $\ket{\Phi}$. (b) Logical view. Each rail denotes an individual frequency bin, with the QFP functioning as a complex interferometer.}
    \label{fig:QFP}
\end{figure}

Up to this point, the mathematical formulation has been completely general with respect to the underlying optical modes, applicable equally well to any photonic DoF. In this section, we refine our focus to frequency bins specifically. Fundamentally, the QFP is designed to realize arbitrary unitary operations on a discrete set of equispaced, clearly separated frequency modes, or bins. Inspired by the LOQC approach of Knill, Laflamme, and Milburn~\cite{Knill2001}---whereby single photons, linear optics, detectors, and feed-forward unite for universal quantum computing---the original QFP proposal~\cite{Lukens2017} succeeded in showing that EOMs~\cite{Wooten2000} and pulse shapers~\cite{Weiner2000, Weiner2011}, alternating in series, could realize a universal gate set, arguing further that any unitary could be synthesized such that the combined number of EOMs and pulse shapers $Q$ (see Fig.~\ref{fig:QFP}) scales like $\mathcal{O}(d)$, where $d$ is the dimension of the targeted unitary.

In order to understand the basic principles of operation, consider a discrete set of frequency modes, each centered at $\omega_n=\omega_0 + n\Delta\omega$ ($n\in\mathbb{Z}$) and associated with an annihilation operator $\hat{a}_n$. The corresponding output operators $\hat{b}_n$ relate to the inputs $\hat{a}_n$ as
\begin{equation}
\label{eq:PS}
\hat{b}_n = e^{i\phi_n} \hat{a}_n
\end{equation}
for a line-by-line pulse shaper and 
\begin{equation}
\label{eq:EOM}
\hat{b}_n = \sum_{k=-\infty}^\infty f_{n-k}\hat{a}_k
\end{equation}
for an EOM driven with phase function $\varphi(t)$ periodic at the inverse mode spacing $T=\frac{2\pi}{\Delta\omega}$, so that $e^{i\varphi(t)} = \sum_n f_n e^{-in\Delta\omega t}$ and $f_n = \frac{1}{T} \int_0^T dt\, e^{i\varphi(t)} e^{in\Delta\omega t}$. As written, this formulation contains an infinite number of frequency bins; in the interests of numerical tractability, though, we can limit the total number of considered modes to $N$ and discretize the temporal period as $t_n = \frac{nT}{N}$ ($n\in\{0,1,...,N-1\}$). Under this approximation, the total $N\times N$ unitary for a sequence of $Q$ components becomes~\cite{Lukens2017}
\begin{equation}
\label{eq:U}
U = (FD_Q F^\dagger) D_{Q-1} \cdots (FD_3 F^\dagger) D_2 (FD_1 F^\dagger),
\end{equation}
where $F$ is the discrete Fourier transform with elements $F_{mn} = \frac{1}{\sqrt{N}} e^{2\pi i mn/N}$ ($m,n \in \{0,1,...,N-1\}$). Each $D_q$ is a diagonal unitary matrix; the odd-numbered $q$ signify an EOM with elements $(D_q)_{nn} = e^{i\varphi^{(q)}(t_n)}$, and the even-numbered $q$ a pulse shaper with $(D_q)_{nn} = e^{i\phi^{(q)}_n}$. We bookend the QFP with EOMs in our example, rather than pulse shapers, based on previous experience where we have observed no increase in circuit performance with the addition of a front- or back-end pulse shaper~\cite{Lu2020b}. 

We note that the alternating pattern in Eq.~\eqref{eq:U} also makes sense conceptually: each device multiplies the input field by a phase-only function, either in the time domain for the EOM or the frequency domain for the pulse shaper; thus, discrete Fourier matrices appear naturally as finite approximations to the continuous Fourier transformations between time and frequency representations. Accordingly, the form in Eq.~(\ref{eq:U}) accurately reflects the physical situation as long as $N$ is sufficiently large so that photon probability amplitudes do not reach the edge of the truncated simulated domain and artificially ``wrap around'' to the other side; in practice, this situation can be avoided by limiting the maximum EOM modulation index or applying bandpass filters to the pulse shaper matrices.

Diagonal unitary decompositions in the form of Eq.~(\ref{eq:U}) have appeared in a variety of photonic DoFs, including position/momentum~\cite{Mueller1998, Morizur2010}, parallel waveguides~\cite{Saygin2020}, and time bins~\cite{Lukens2018}---whenever the physical system can be modeled as the application of phase shifts in alternating Fourier-transform pairs. As shown in Ref.~\cite{Lopez2019}, one can analytically design such systems by starting with the beamsplitter/phase-shifter decomposition of path encoding~\cite{Reck1994, Clements2016}, and then expressing each beamsplitter layer as six alternating phase masks; however, this introduces significant resource overhead, so that there currently exists no recipe to compute the $D_q$ matrix elements required to synthesize a desired target matrix $U$ \emph{optimally}---i.e., without an intermediate conversion step to an equivalent path circuit. Accordingly, numerical optimization has been employed extensively in QFP designs for basic gates such as the Hadamard~\cite{Lu2018a, Lu2018b}, controlled-NOT~\cite{Lu2019a}, cyclic hop~\cite{Lu2020a}, and arbitrary single-qubit unitaries~\cite{Lu2020b}. From the perspective of photon statistics, the most complicated QFP gate explored so far is the two-ancilla controlled-$Z$ in Ref.~\cite{Lukens2017} containing a total of four photons. 
In contrast, the non-Gaussian CV cases considered in the present work deal with many-photon states inherently, so the mathematics involved proves markedly more complex.

Figure~\ref{fig:QFP} provides an overview of our non-Gaussian state engineering system. As previously mentioned in Sec.~\ref{sec:Mathematical Background of our Approach}, our mathematical formalism applies to a system like Fig.~\ref{fig:GenericScheme} where the $M$ undetected modes can be any of the total $N$ modes without loss of generality. In the following application we choose $M = 1$ and select this single undetected mode as the $K$th mode, which is at the center of a set of $N_{s}$ modes that are populated with single-mode squeezed vacuum states at the input; the remaining $N-N_s$ modes are initially vacuum. For our simulations, we take the phase of the squeezing to be zero for all cases. After application of $U$, the $K$th mode is left undetected and $n_{s}$ photons are detected in each of the other central $N_s - 1$ modes. Production of the desired state in the $K$th mode is heralded by simultaneously detecting vacuum in the remaining $N - N_s$ modes: in essence, a bucket detector for all remaining modes, reminiscent of heralded QFP gates in the DV case~\cite{Lukens2017}. The requirement of such vacuum postselection is a consequence of the presence of an infinite set of ancilla modes in the frequency-bin DoF, which must be detected to ensure that the output state is pure. 

Given the massive design space available for non-Gaussian QFP circuits---in terms of input states, unitaries, and output patterns---we have attempted in the specific configuration of Fig.~\ref{fig:QFP} to provide a relatively simple construction that nevertheless retains key features anticipated for successful circuits. By placing the output mode of interest in the center of the squeezed inputs, we maximize opportunities for multiphoton interference with relatively weak modulation amplitudes, and selecting from the modes initially populated with photons for PNR detection with $n_s>0$ should permit reasonable success probabilities.  Of course, there is no guarantee that such intuitions are globally optimal, and thus work into other configurations will be extremely valuable in the future such as, e.g., including the choice of detection pattern within the optimization algorithm itself, rather than taking it as given. Yet the present setup offers a feasible foundation for this initial investigation.

Finally, before proceeding further, we note that several features of the design in Fig.~\ref{fig:QFP} share commonalities with previous work in frequency-based quantum information. Extensive research in CV quantum frequency combs~\cite{Pysher2011, Chen2014, Pfister2020, Yang2021} has focused on frequency-bin encoding for cluster-state--based quantum computing; in fact, a recent theoretical investigation specifically incorporated EOMs, finding that highly intricate, multidimensional cluster states could be produced by modulating a comb of two-mode squeezed states at multiples of the frequency-bin spacing~\cite{zhu2021hypercubic}. Although these aspects overlap strongly with our approach, the addition of pulse shaper layers in the QFP  provides considerably more complexity in the unitaries available, and our explicit examination of PNR detection allows us to attain non-Gaussian states that are not available within existing Gaussian cluster state models. On another front, researchers have recently introduced an alternative GKP qubit encoding consisting of a single photon in a discrete grid of spectro-temporal modes~\cite{Fabre2020}. While similar in that this also leverages the frequency DoF, we follow the more traditional construction of non-Gaussian states in which quantum information resides in the field quadratures of optical modes, making our analysis inherently multi- rather than single-photon in nature.


\section{Circuit Design Examples}
Having detailed the mathematical formalism in Sec.~\ref{sec:Mathematical Background of our Approach} and highlighted the specific features of the QFP in Sec.~\ref{sec:QFP}, we now apply the complete framework toward the design of quantum circuits that produce desired non-Gaussian output states, according to the configuration presented in Fig.~\ref{fig:QFP}.

\subsection{Numerical optimization}\label{sec:numOpt}
As noted in the previous section, quantum system design with the QFP lacks an optimal analytical unitary decomposition procedure, so that numerical optimization is in general required to obtain a QFP configuration realizing a desired unitary. In the context of non-Gaussian state design, the need for numerical optimization in itself is not unique, but has proven a 
fixture in path encoding as well
~\cite{Sabapathy2019, Su2019, Tzitrin2020}. However, the QFP case does present additional practical constraints, most notably with respect to the stellar decomposition~\cite{Chabaud2020} leveraged in previous path-encoded  designs~\cite{Sabapathy2019, Su2019}.

In this approach, rather than designing a quantum circuit to implement some target state $\ket{\Phi_t}$ directly, a Fock-truncated core state $\ket{\Phi_\mathrm{core}}$ is sought instead---related to $\ket{\Phi_t}$ via a squeezing and displacement operation, $\ket{\Phi_t} = S(z)D(\beta)\ket{\Phi_\mathrm{core}}$. Suppose that the mode unitary found to produce $\ket{\Phi_\mathrm{core}}$ is $U$; then, by absorbing the displacement and squeezing operation into a new set of inputs and mode unitary $U^\prime$, an interferometer for the desired full state $\ket{\Phi_t}$ can be produced immediately via the analytical decomposition scheme of Refs.~\cite{Reck1994, Clements2016}. In the QFP case, however, if a set of EOM and pulse shaper solutions are found that can implement the core state preparation circuit $U$, the absence of an available analytical decomposition procedure means that there is no functional connection from this solution [i.e., the $D_q$ matrices in Eq.~(\ref{eq:U})] to the modified configuration that would realize $U^\prime$; instead, numerical optimization must again be employed on $U^\prime$, effectively doubling the rounds of numerical design compared to path encoding. 
Accordingly, in what follows we concentrate on synthesizing circuits that produce the full target state $\ket{\Phi_t}$ immediately, avoiding this intermediate core state step.

To begin the optimization process we first define the target state $\ket{\Phi_t}$ in the Fock basis, i.e., the coefficients $\tau_n = \braket{n|\Phi_t}$. We employ MATLAB's particle swarm optimization (PSO) tool~\cite{PSO} to find the $N_s$ nonzero input squeezing values of the total length-$N$ vector of inputs
\begin{multline}
\label{eq:r}
\vec{r} = (0,\ldots,0,r_{K -\lfloor\frac{N_s}{2}\rfloor},\ldots,r_{K-1},r_K, \\ r_{K+1},\ldots,r_{K +\lfloor\frac{N_s}{2}\rfloor},0,\dots,0),
\end{multline}
and a QFP unitary, $U$, that when applied to the $N$-mode input followed by detection of $n_s$ photons in each of the remaining $N_s-1$ squeezed modes, produces a state $\ket{\Phi}$. Letting $n_c$ denote the photon number at which we truncate the state for numerical simulations, we therefore must compute a total of $n_c+1$ coefficients (including vacuum) to fully describe the heralded output.
To find the optimal squeezing values and $U$, PSO varies the phase shifts applied to the $N$ QFP modes by the pulse shapers, each EOM's phase modulation function $\varphi(t)$, and the $N_s$ nonzero elements of $\vec{r}$ in order to minimize the cost function 
\begin{equation}
    \mathcal{C} = P \log_{10}(1 - \mathcal{F}),
    \label{eq:costfunc}
\end{equation}
where $\mathcal{F}$ and $\textit{P}$ are the fidelity of $\ket{\Phi}$ with respect to $\ket{\Phi_t}$ and the probability of producing $\ket{\Phi}$, respectively. We have found a logarithmic cost function of this form useful for penalizing $\mathcal{F}<1$ more strongly than $\textit{P}<1$, emulating the effect of a constraint on $\mathcal{F}$ without the computational cost associated with a strict constraint function. With the revelations of Secs.~\ref{sec:Gamma} and \ref{sec:H} the Fock coefficients of $\ket{\Phi}$ in the $K$th mode can be expressed as Eq.~\eqref{eq:FockCoef} which we write in the form 
\begin{equation}
\label{eq:Fockcoeff_N}
c_{n_K} = \frac{\cI_{\vec{n}}}{\sqrt{P} \prod\limits_{i=1}^{N} \sqrt{n_i!2^{n_i} \cosh r_i}},
\end{equation}
where $\vec{n}=(0,...,0,n_s,...,n_s,n_K,n_s,...,n_s,0,...,0)$ is the vector of photon numbers over all output modes, so that $P$ and $\mathcal{F}$ can be written as 
\begin{equation}
\label{eq:prob} P = \sum_{n_K=0}^{n_c} \left( \frac{|{\cI_{\vec{n}}}|^2}{ \prod\limits_{i=1}^{N} n_i!2^{n_i} \cosh r_i} \right)
\end{equation}
and
\begin{equation}
\label{eq:fid}	
\mathcal{F} = |\langle\Phi_{t}|\Phi\rangle|^{2} = \left| \sum_{n_K=0}^{n_c} \tau_{n_K}^* c_{n_K} \right|^2
\end{equation}

With the cost function defined we now lay out the recipe for evaluating $\mathcal{F}$ and $P$ at each PSO iteration. First, 
we calculate $(V_{0} + \frac{I}{2})^{-1}$ using $\vec{r}$ in Eq.~\eqref{eq:invGamma_0}. $U$ is calculated by substituting the $N$ phase shifts for each pulse shaper and each EOM's $\varphi(t)$ into Eq.~\eqref{eq:U}, which we convert to symplectic form, $S_{p}$, via 
\begin{equation}
\label{eq:UtoS_p}S_{p} = W
\begin{pmatrix}
U & \mathbf{0}\\
\mathbf{0} & U^{*}
\end{pmatrix}W^{\dagger},
\end{equation}
where $W$ is defined by Eq.~\eqref{eq:Wmatrix} and $U^*$ corresponds to element-by-element conjugation (no transpose). $\Gamma^{-1}$ is then calculated by Eq.~\eqref{eq:GammaInv2}, and the blocks $A$ and $C$ extracted per Eq.~\eqref{eq:GammaInv3}. $A$ and $C$ are used to find $\mathcal{H}^{-1}$ with Eq.~\eqref{eq:Hinv}. The matrix elements of $\sigma$ are found by using $\mathcal{H}^{-1}$ in Eq.~\eqref{eq:FandHinv2}. Because we detect vacuum in all the QFP modes except for the center $N_s$, $\vec{h}^{T} = (0,\ldots,0,\frac{n_s}{2}-\nu_{K-\lfloor\frac{N_s}{2}\rfloor},\ldots,\frac{n_K}{2}-\nu_{K},\ldots,\frac{n_s}{2}+\nu_{K+\lfloor\frac{N_s}{2}\rfloor},0,\ldots,0)$ in Eq.~\eqref{eq:FasterWick} renders  unimportant all the elements of $\sigma$ other than the center $N_s\times N_s$ block. Therefore we proceed to evaluate Eq.~\eqref{eq:Fockcoeff_N} using only the center block of $\sigma$, for $n_{K} \in \{0,1, \ldots, n_{c}\}$ and are left with the Fock coefficients $c_{n_K}$ of $\ket{\Phi}$. 

We choose to compute $\Gamma^{-1}$ in this manner for computational reasons.
As a large matrix---$2N\times 2N$ in general and $128\times 128$ in our case---$\Gamma$ is time-consuming to invert. 
We bypass this time sink by calculating $\Gamma^{-1}$ directly with Eq.~\eqref{eq:GammaInv2}, rather than performing $\Gamma = S_{p}V_{0}S_{p}^{T} + I/2$ and inverting $\Gamma$. An alternative route to reaching $\Gamma^{-1}$ is to calculate the $A$ and $C$ matrices using Eqs.~\eqref{eq:blockA} and \eqref{eq:blockC}, respectively, and then substituting them into Eq.~\eqref{eq:GammaInv3}. While valid, this path involves four separate matrix products, making it less computationally efficient than using Eq.~\eqref{eq:GammaInv2} that requires only one matrix product.

We take further action to streamline the $n_{c} + 1$ calculations of $\cI_{\vec{n}}$ needed to find the Fock coefficients of $\ket{\Phi}$, dominated by $\textrm{Hf}(\sigma)$ in Eqs.~\eqref{eq:Hafnian} and \eqref{eq:FasterWick}. $\sigma$ is the only quantity in Eq.~\eqref{eq:FasterWick} that will change in the successive iterations of PSO; therefore, we can precompute a number of the elements of Eq.~\eqref{eq:FasterWick} outside the optimization loop and use them for every PSO iteration. Calculating these elements upfront proves imperative to expediting the optimization process when $n_c$ becomes large. 
 
Consider a given value $n_K$. First, we define the length-$N_s$ vector
\begin{equation}
\label{eq:s}
\vec{s}^T = (n_s,...,n_s,n_{K},n_s,...,n_s),
\end{equation}
the sum of photons in the output modes
\begin{equation}
\label{eq:sum}
\Sigma=\sum_{i=1}^{N_s} s_i = (N_s-1)n_s + n_K,
\end{equation}
and index vectors for each mode $\vec{\nu}_i^T = (0,1,...,s_i)$, where the maximum $s_i$ for each mode is taken from Eq.~\eqref{eq:s}.
 
Then we find all combinations of the entries of the $\nu$ vectors and store them in a matrix $D$, where each row corresponds to a unique length-$N_s$ listing of elements, one drawn from each $\vec{\nu}_i$. Because we choose to detect the same number of photons, $n_s$, in the $N_s -1$ modes, $D$ will be of dimension $(n_s +1)^{N_s -1}(n_K +1) \times N_s$ and will take on the role of the nested summations that appear in Eq.~\eqref{eq:FasterWick}. We calculate the exponent of the $(-1)$ factor in Eq.~\eqref{eq:FasterWick} for all terms of the nested summation and store them in $\vec{\mathcal{W}}$ whose elements are defined as 
\begin{equation}
\mathcal{W}_{i} = \sum\limits_{j=1}^{N_s}{D_{ij}}.
\end{equation}

Similarly, the product of binomials in Eq.~\eqref{eq:FasterWick} is calculated for all terms in the nested summation and stored in $\vec{\mathcal{X}}$,  
\begin{equation}
\mathcal{X}_{i} = \prod\limits_{j=1}^{N_s}\binom{s_j}{D_{ij}}.
\end{equation}
$\vec{h}^{T}$ for all the terms in the nested summation are stored in vectors 
\begin{equation}
\vec{\mathcal{Z}}_{i}^{T} = \left(\frac{s_1}{2}-D_{i1},~\ldots~, \frac{s_{N_s}}{2}-D_{iN_s}\right).
\end{equation}
The $\textrm{Hf}(\sigma)$ calculation is then reduced to a single summation over these precomputed elements,
\begin{equation}
 \textrm{Hf}(\sigma) = \sum\limits_{i = 1}^{\kappa} \frac{1}{\left(\frac{\Sigma}{2}\right)!} (-1)^{\mathcal{W}_{i}} \mathcal{X}_{i} \left(\frac{1}{2}\vec{\mathcal{Z}_{i}}^T \sigma \vec{\mathcal{Z}_{i}} \right)^\frac{\Sigma}{2},
\end{equation}
where $\kappa = (n_s+1)^{N_s -1}(n_K +1)$. Keep in mind this process must be repeated for all values $n_{K} \in \{0,1, \ldots, n_{c}\}$ giving us 
all necessary precomputed $\vec{s}$, $\Sigma$, 
$\vec{\mathcal{W}}$, $\vec{\mathcal{X}}$, and $\vec{\mathcal{Z}}_i$ elements.

\subsection{Coherent cat states}\label{sec:Coherent Cat States}
As examples of our method, we seek to generate even Schr\"{o}dinger cat states with coherent amplitudes $\alpha$ ranging from 0.5 to 3 in steps of 0.25. For each $\alpha$ value $\ket{\Phi_t}$ is therefore set to
\begin{equation}
\label{eq:EvenCat}
    \ket{\Phi_t} = \frac{\ket{\alpha} + \ket{-\alpha}}{\sqrt{2(1+e^{-2|\alpha|^{2}})}},
\end{equation}
where $\ket{\pm \alpha} \approx e^{-\frac{1}{2}|\pm \alpha|^{2}}\sum_{n = 0}^{n_{c}}\frac{(\pm \alpha)^{n}}{\sqrt{n!}}\ket{n}$. We truncate $\ket{\pm \alpha}$ at $n_{c} = 40$ for all $\alpha$ values, which encompasses all the Fock support to high precision at $\alpha = 3$, and therefore for any $\ket{\Phi_{t}}$ with $\alpha < 3$ as well. Indeed, the truncation error defined as $\epsilon_{n_c} = 1-\sum_{n=0}^{nc} |\braket{n|\Phi_{t}}|^{2}$ is less than $10^{-14}$ for $\alpha\leq 3$ and $n_c=40$. This choice is highly conservative, as one could likely consider smaller $n_c$ values such as $n_{c} = 20$ or 30 for added computational speed up, for which the errors remain small: $\epsilon_{20} < 10^{-3}$ and $\epsilon_{30} <10^{-8}$ at $\alpha = 3$.   

To make these results as tractable as possible for experiment we limit $\varphi(t)$ to a single sinewave and constrain $(r_{K -\lfloor\frac{N_s}{2}\rfloor},\ldots,r_{K},\ldots,r_{K +\lfloor\frac{N_s}{2}\rfloor})$ to a maximum of 1.5 (corresponding to a squeezing value of approximately 13~dB). We proceed to optimize with $Q \in \{3,5,7\}$ total QFP elements, $N=64$ QFP modes, and $N_s \in \{3,5\}$ input squeezed states, along with a 32-mode bandpass filter on each pulse shaper to prevent unphysical solutions that reach the edge of the $N=64$-mode truncation. The nonzero PNR detectors are set to herald on $n_s = 1$, which ensures that $\Sigma$ in Eq.~\eqref{eq:Hafnian} will be even when computing even Fock coefficients in the undetected mode $K$ ($n_{K} \in \{0,2,\ldots,40\}$). 
The target cat state coefficients are real numbers; however, the coefficients found by optimization are in general 
complex. Therefore if the state found by optimization is perfect (fidelity equal to one), it should have a constant phase for all Fock coefficients. 

\begin{figure*}[!ht]
    \centering
    \includegraphics[width=\textwidth]{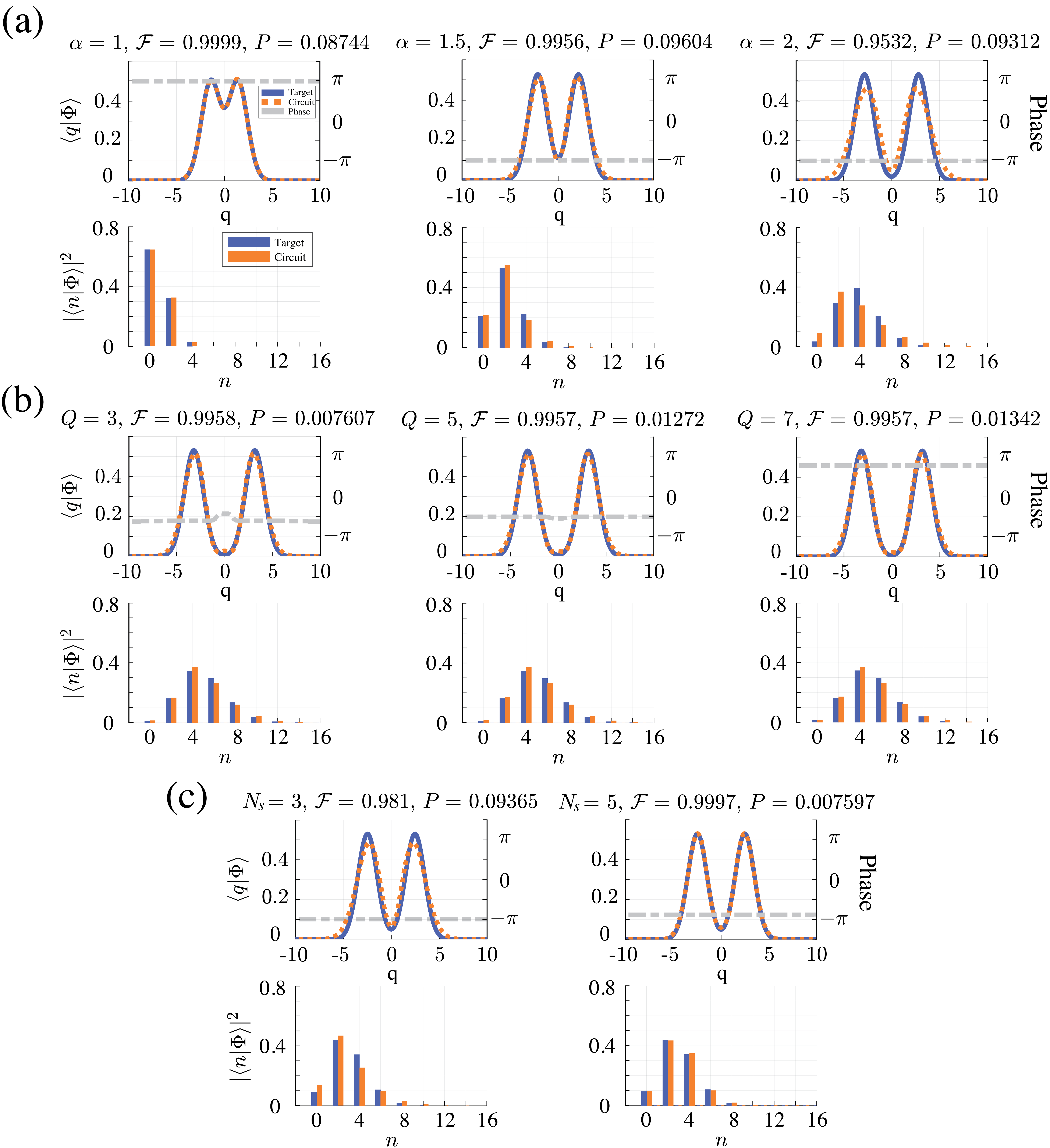}
    \caption{Wavefunctions in the quadrature basis $\langle \textit{q}| \Phi \rangle$ (top) and photon number probabilities $|\langle \textit{n}| \Phi \rangle|^{2}$ (bottom) for example target and QFP output states. (a)~$\alpha \in\{1,1.5,2\}$, $Q=3$, and $N_s=3$. (b) $\alpha=2.25$, $Q \in \{3,5,7\}$, and $N_s=5$. (c) $\alpha=1.75$, $Q=3$, and $N_s \in \{3,5\}$.
    }
    \label{q_plots_all}
\end{figure*}

\begin{figure*}[!ht]
    \centering
    \includegraphics[width=\textwidth]{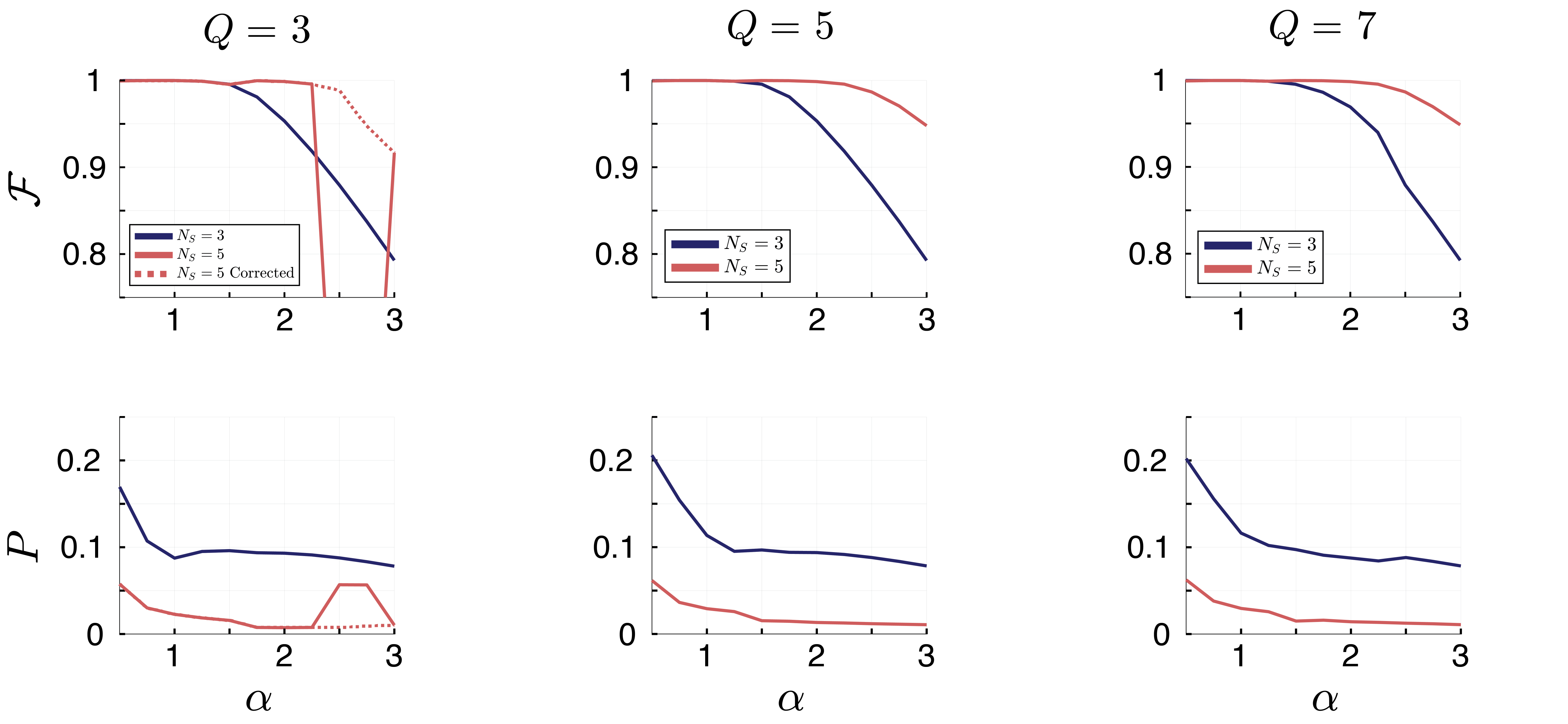}
    \caption{Fidelity (top) and Success Probability (bottom) plotted as functions of $\alpha$ (in steps of 0.25) for various combinations of $N_s$ and $Q$. Corrected results are shown for the $N_s = 5, Q = 3$ case (see Sec.~\ref{sec:Coherent Cat States} for disambiguation). }
    \label{F&P}
\end{figure*}

To elucidate how the size of the cat state changes with $\alpha$ we plot, in Fig.~\ref{q_plots_all}(a), $\ket{\Phi_t}$ (target) and $\ket{\Phi}$ (circuit), with $N_s = 3$ and $Q = 3$, for $\alpha \in\{1,1.5,2\}$. The plots in Fig.~\ref{q_plots_all}(b) and~\ref{q_plots_all}(c) illustrate how the quality of $\ket{\Phi}$ changes with $N_s$ and $Q$ for single $\alpha$ values, whereas Fig.~\ref{F&P} shows the overall trends. While running PSO it became apparent that our chosen cost function [Eq.~\eqref{eq:costfunc}] did not favor high-fidelity solutions as strongly as intended, but in certain cases converged to solutions with higher $P$ but $\mathcal{F} \ll 1$. For example, in Fig.~\ref{F&P} for $\alpha = 2.5$, $Q=3$, and $N_s = 5$, the output $\ket{\Phi}$ with the lowest $\mathcal{C}$ is a state with $\mathcal{F} = 0.47$ (not even visible in the plotted range). Consequently, we include in Fig.~\ref{F&P} not only the states $\ket{\Phi}$ with the lowest cost $\mathcal{C}$, but also higher cost solutions, found with different initial conditions, that attain fidelities $\mathcal{F}>0.9$ (corrected). We emphasize that this distinction does not reflect any issues in the optimization procedure itself, but rather in our selection of the cost function; to encourage PSO to find even higher fidelity states, future tests could consider alternative cost functions that more aggressively penalize low fidelities.

Our results are comparable to those achieved by a similar photon subtraction method performed in the path DoF by Quesada \textit{et al.}~\cite{Quesada2019}: for an even cat state with $\alpha \approx 1.3$ and zero loss, both approaches produce states with similar fidelity. Our states do exhibit a higher success probability; however, this improvement is expected as Ref.~\cite{Quesada2019} uses a single squeezed input with a fixed value while we optimize our $N_s$ individual input squeezing values. And although the impact of probabilistic state production will depend on both the protocol implemented and available resources, we nevertheless note that the range of values found here ($0.01\lesssim P \lesssim 0.2$) are of the same order as many standard gates in DV LOQC with unentangled ancillas---e.g., the heralded controlled-NOT succeeds with $P=2/27$~\cite{Uskov2009})---suggesting that they are in a reasonable scale for photonic quantum information processing.

For a set amount of resources, constant $N_s$ and $Q$, the output state quality found by PSO finds decreases as $\alpha$ increases. This can be attributed to the fact that $\ket{\Phi_t}$ becomes noticeably more non-Gaussian as $\alpha$ is increased [see Fig.~\ref{q_plots_all}(a)]. 
Figure~\ref{F&P} reveals that while increasing the complexity of the QFP through the number of elements $Q$ can moderately improve the success probability [cf. Fig.~\ref{q_plots_all}(b)], it does not lead to markedly higher fidelities in these examples. In contrast, for any $Q$, the addition of more ancilla resource states (larger $N_s$) can substantially improve fidelity, particularly for larger values of $\alpha$, albeit with about an order of magnitude reduction in success probability [see Fig.~\ref{F&P}].
Intuitively, this behavior makes sense; the extra photons available provide a greater variety of interference possibilities in design, yet also reduce the success probability through additional PNR detector conditions that must be satisfied.

\section{Discussion}\label{sec:Disc}
\subsection{Further generalizations}
The formalism we have presented currently relies on several assumptions, most notably specialization to single-mode squeezed vacuum inputs and the neglect of photon loss. Mathematically speaking, single-mode squeezed vacuum states are especially convenient because of their zero displacement in phase space ($\vec{x}_\beta=0$) and diagonal covariance matrix $V_0$. Importantly, the latter facilitated closed-form expressions for $\det\Gamma$ [Eq.~\eqref{eq:detGamma2}] and $\Gamma^{-1}$ [Eq.~\eqref{eq:GammaInv3}], markedly simplifying calculations for the numerical optimizer. However, the covariance matrix just before the partial PNR detection (i.e., just after the passive Gaussian unitary operator $U$ of Fig.~\ref{fig:GenericScheme}) nevertheless remains completely general for a pure Gaussian state. Indeed, the covariance matrix of a pure Gaussian state is of the form $S V_{\text{vac}} S^T$, where $V_{\text{vac}}$ is the covariance matrix of vacuum and $S$ is any symplectic matrix which includes squeezing, i.e., $S=S_p S_s S_p'$, where $S_p$ and $S_p'$ are passive transformations and $S_s$ is the symplectic transformation for squeezing. Since passive transformations have no effect on vacuum, the most general covariance matrix for a pure Gaussian state can thus be written as $S V_{\text{vac}} S^T = S_p V_0 S_p^T$, which is precisely the covariance matrix assumed in our analysis. 
For example, any two-mode squeezed state (like those produced in quantum frequency combs~\cite{Pysher2011, Chen2014, Pfister2020, Yang2021}) can be expressed as the interference of two single-mode squeezed states on a beamsplitter, whose unitary can be readily incorporated on the front-end of the circuit in Fig.~\ref{fig:QFP}. 

Yet although the diagonal input covariance matrix $V_0$ does not reduce the generality of our formulation, the absence of displacement is significant.
Incorporating nonzero displacements will not affect the covariance matrix we have used; it will, however, introduce additional variables into the optimization procedure for generating desired non-Gaussian states. Since we have been able to obtain high fidelities for our purposes using zero-displacement inputs only, we leave the effects of displacement to be thoroughly studied in the future.

It should be possible to move beyond unitary operations as well. For example, by coupling each frequency bin to additional environmental modes, then tracing these out, photon loss can be added into the formulation, following the outline in Ref.~\cite{Gagatsos2019}. The specifics of how the final expressions can be simplified in this case---as well as how they might compare with those of the $Q$ function approach adopted for loss in Ref.~\cite{Quesada2019}---remain open questions. Nevertheless, such an extension will be extremely important from an experimental perspective, and new in the context of QFP design. QFP theory up to this point has concentrated on DV gates with Fock states, where loss reduces photodetection events but does not otherwise modify the (postselected) quantum state. On the other hand, the prepared states here depend heavily on both loss and detector efficiency, making this elaboration critical to predicting experimental performance. Moreover, in light of the insertion loss of commercial discrete fiber-pigtailed EOMs and pulse shapers, both loss modeling and loss mitigation will be vital in advancing this field. To this end, integrated EOMs~\cite{Wang2018b, Ren2019} and pulse shapers~\cite{Khan2010, Wang2015b} with the potential for much higher efficiencies seem particularly promising, and in our view on-chip QFP integration is a prerequisite for practicable non-Gaussian state generation according to the approach proposed here.

\subsection{GKP states}
While the generation of cat states is nontrivial in itself, a long-standing challenge in CV encoding is the realization of GKP qubit states for error correction. 
The value of GKP qubits, call them $\ket{0}$ and $\ket{1}$ in the logical basis, lies in their infinite series of equispaced delta functions, $\ket{1}$ being displaced from $\ket{0}$ by $\sqrt{\pi}$ when plotted in the \textit{q}-quadrature basis. Since these ideal states are unphysical, approximate states $\ket{\tilde{0}}$ and $\ket{\tilde{1}}$ were presented in the original GKP proposal~\cite{Gottesman2001}, which consist of a sum of Gaussian peaks with standard deviation $\Delta$, all under another Gaussian envelope with standard deviation $\frac{1}{k}$. $\Delta = k = 0.15$ is required for $\ket{\tilde{0}}$ and $\ket{\tilde{1}}$ to maintain a 99\% error correction rate~\cite{glancy2006error}. Due to the limited understanding of how $n_{s}$ and ancilla mode placement affect the quality of the output state, finding effective QFP circuits for direct GKP state production is beyond the scope of the present investigation,
but provides an important direction for future work.

An alternative path to quality approximations of GKP qubits, for which our system is already well suited, is the so-called ``cat breeding'' protocol~\cite{vasconcelos2010all, weigand2018generating, eaton2019non}.
In the first version of the protocol~\cite{vasconcelos2010all}, two cat states are squeezed by some amount $r$, where $r = -\ln\Delta$. The squeezed cat states are combined on a balanced beamsplitter, and a homodyne measurement is made on one of the output modes. When the result of the homodyne measurement for a single output mode's \textit{p}-quadrature is zero, the other output mode is left in a state with three equispaced peaks. The height of the peaks follows a binomial distribution, and the width of the peaks is determined by the amount of squeezing applied to the initial cat states. Successive iterations of the protocol, where the beamsplitter inputs are the states produced by the previous iteration, yield higher-order binomial states. To ensure that the final state has the correct spacing associated with GKP states, the initial cat states must have a coherent amplitude $\alpha = \sqrt{2}^{m-1}\sqrt{\pi} e^{r}$, where $m$ is the number of iterations of the protocol to be executed. The larger $r$ and $m$ are, the more closely the resulting state will resemble the approximate GKP state, making access to large cat states vital to the protocol. The version of the protocol presented by Eaton~\textit{et al.}~\cite{eaton2019non} replaces the homodyne measurement by PNR detection. Because PNR detection neglects the phase of the output, fine control over the relative phase of the input states is needed to achieve the same comb-like output as in the homodyne approach. By detecting four photons at one output mode after a single iteration of the protocol, Ref.~\cite{eaton2019non} numerically generated states with a fidelity of 0.996 with respect to an approximate GKP state ($\Delta = k = 0.545$) at a success probability of 0.09.

As presented in Sec.~\ref{sec:Coherent Cat States} our system can generate cat states up to a size $\alpha = 2$ with $99.87$\% fidelity when $N_s = 5$ and $Q = 7$. These capabilities make our non-Gaussian state engineering system a viable candidate to meet the resource state demands set by cat breeding protocols.

\section{Conclusion}\label{sec:Conclusions}
We have introduced a complete model for the production of non-Gaussian quantum states using the QFP, a device designed to implement arbitrary linear-optic transformations on discrete spectral modes. Our mathematical formulation using the $K$ function expansion enables efficient calculation of multimode Gaussian states in the photon-number basis, providing a valuable framework for analysis in any photonic DoF. Applying this to the QFP specifically, we have designed basic quantum circuits that produce non-Gaussian cat states with a variety of amplitudes, revealing a clear fidelity/success-probability tradeoff with the number of squeezed ancillas. Given the multitude of configurations possible---along with the rapidly evolving nature of quantum computation with non-Gaussian resources such as GKP qubits---many unsolved challenges remain on the path toward large-scale quantum information processing in this paradigm. Nonetheless, our work furnishes an important foundational tool for designing CV quantum systems in frequency bins and should contribute toward the realization of fiber-compatible, single-spatial-mode, and parallelizable quantum information processors based on non-Gaussian photonic states.

\section*{Data Availability}
Data and MATLAB codes used in this paper are available from A.J.P. on request (ajpizzimenti@email.arizona.edu).

\acknowledgments
We thank R.~C. Pooser and K.~K. Sabapathy for useful discussions. This research was performed in part at Oak Ridge National Laboratory, managed by UT-Battelle, LLC, for the U.S. Department of Energy under contract no. DE-AC05-00OR22725. Funding was provided by the U.S. Department of Energy, Office of Science, Office of Advanced Scientific Computing Research, through the Transparent Optical Quantum Networks for Distributed Science Program and Early Career Research Program (Field Work Proposals ERKJ355 and ERKJ353). A.J.P. acknowledges support from the U.S. Department of Energy, Office of Science, Office of Workforce Development for Teachers and Scientists Science Undergraduate Laboratory Internship Program. C. N. G. and S. G. acknowledge an ORNL/DOE subaward under grant number 4000178321, and partial support from the Office of Naval Research (ONR) under grant number N00014-19-1-2189.

\bibliography{BIB.bib}
\end{document}